\documentclass[review]{elsarticle}
\usepackage{amsmath}
\usepackage{geometry}
\usepackage[varg]{txfonts}
\usepackage{graphicx}
\usepackage{natbib}
\usepackage{pifont}
\usepackage{xcolor}
\biboptions{authoryear} 
\usepackage{hyperref}

\begin{document}

\title{Orbit dissimilarity criteria in meteor showers: a comparative review}

\author[1,2]{Ariane Courtot \corref{cor1}}
\ead{ariane.courtot@obspm.fr}
\author[2,3]{Patrick Shober}
\author[2]{Jérémie Vaubaillon}

\cortext[cor1]{Corresponding author}
\affiliation[1]{organization={ESA Space Environments and Effects Section (TEC-EPS), ESTEC},
city = {Noordwijk},
postcode = {2201AZ},
country = {The Netherlands}}
\affiliation[2]{organization = {LTE, Observatoire de Paris, PSL Research University, Sorbonne Universit\'{e}, Université de Lille, LNE, CNRS}, 
city = {Paris},
postcode = {75014},
country = {France}}
\affiliation[3]{organization={NASA Astromaterials Research and Exploration Science Division, Johnson Space Center}, 
city = {Houston},
postcode = {TX 77058}, 
country = {USA}}

\date{Received 25 July 2025 / Accepted 9 December 2025}

\begin{abstract}
In meteor science, the identification of meteor showers is a crucial and complex problem. The most common method is to perform a systematic search of a database of observed orbits using an orbit dissimilarity criterion (D-criterion) and an algorithm. D-criteria compare the result of an orbit dissimilarity function (D-function) and a threshold. These D-functions associate one value to two orbits. If this value is lower than the threshold, the orbits are considered similar. In this paper, we focus on the application of these D-criteria on meteoroid orbits. Group of meteors are thus formed using this method. However, not all D-criteria have been evaluated, and their high number makes it hard to know which should be prioritised. This paper presents a review of each D-function, the tests they passed, the threshold choice, and the algorithms they are used with. The aim is both to clearly present the state of the art on this question but also to analyse what studies are missing on this topic. We show what methods are currently used in the search for meteor showers, presenting statistics based on papers justifying the existence of established meteor showers. This paper presents a review of each D-functions from eight different papers. We describe how thresholds are usually chosen and what clustering algorithms can be used with D-criteria to form meteor groups. We also analyse tests that were performed on D-criteria, showing which results they were able to achieve and where they fell short. We discover that most of these criteria were not properly tested, and that some have been criticised for their theoretical background. Thus, we recommend performing a post-search analysis of the groups found, both in a statistical sense (to make sure the groups formed could not have been formed randomly) and an orbital dynamics sense (to check whether the group could indeed come from a singular parent body), to present the findings as potential meteor showers.
\end{abstract}


\maketitle

\section{Introduction}

Meteoroids are small natural objects ejected from a comet or, more rarely, an asteroid. When a meteoroid enters the Earth atmosphere, it becomes detectable as a meteor. We will use the term "meteor shower" when discussing a set of meteors originating from a single parent body. However, we will use the term "meteor group" to talk about a set of meteors that originate from similar orbits. This term is not currently used in the literature, but there are two precedents: we have used it in previous works \citep{Courtot2023} and \citet{Kashcheyev_Lebedinets_1967} use a similar notion. It will be necessary throughout this paper, as it shows clearly the distinction between a "meteor group" and a "meteor shower", the latter adding to the condition of orbital similarity a similar origin, which cannot be proven solely with orbital similarity. Indeed, a statistical analysis should be performed on the groups to prove that the similarity is not due to chance, as well as a dynamical analysis to prove that each meteor have a common origin, even if the parent body does not exist anymore. We will therefore talk about "a meteor group that was found", instead of "a meteor shower that was found", when relevant, and regardless of the words used in the papers cited.

To find meteor showers, many authors start with analysing the radiants of the meteors, as well as the date of detection (or the solar longitude of the meteor). Using these indicators, they are able to find sets of meteors that originated from a similar radiant at a similar time. Then they usually perform other analysis, such as the D-criteria method we will present in the next paragraphs. However, this is not always the case, and the D-criteria method can be enough to form meteor groups, without necessarily discussing the radiant. For example, \citet{Andreic_al_2013,Andreic_al_2014} do not use radiants in their search.

The D-criteria method is pretty straightforward: meteoroids orbits from a database are compared (usually in pair) thanks to an orbit dissimilarity function (D-function, sometimes called abusively D-criterion). When the value given by this function is lower than a predetermined threshold, the meteoroids compared are grouped. When the whole database has been tested, one or more meteor groups are found. A variant of this method can be used to find new objects belonging to an already known shower. In this case, meteoroids orbits will be compared to an orbit characteristic of this shower, such as the mean or median orbit.

The combination of a D-function and its threshold is called a D-criterion. The threshold is thus the maximum value of dissimilarity at which two objects can still be considered to have similar orbits in the context of meteor showers. It should be noted that D-functions have been used for other objects, such as asteroids. However, this paper will focus on the applications of such functions and their thresholds to meteor showers specifically. Therefore we will not discuss any aspects of D-criteria based on asteroid orbits.

The D-criterion method to find new meteor showers, described above, was criticised several times. First, the accuracy of many D-criteria is not clearly established. In other words, while D-criteria provide a group of meteors, it is not always clear if this group could also have been formed randomly. A statistical analysis should, therefore, be performed when a D-criterion is used to make sure the resulting group is relevant. This is especially true because, with the betterment of instruments, the amount of meteoroid detected is growing. The chance of finding several orbits that are similar is thus growing as well.

The second problem is the choice of the D-function, as there are many of them, some with slightly different definitions and others using completely different reasoning. There is currently no complete study of all D-function, comparing them together and pointing out which ones work best in what situation, which means the community usually uses D-functions that are well-known, even if they are not proven to be the most reliable. The first D-function defined by \citet{Southworth_Hawkins_1963} is the most commonly used, despite several major criticisms against it. One of the aims of this paper is to provide the reader with other criteria to choose from.

The choice of the threshold is a difficulty in itself. Several methods have been developed over the years to make this choice as reliable as possible, as we will discuss in this paper.

A clustering algorithm is also needed to form meteor groups from the database studied and using the D-criterion. Once more, there are several to choose from, each with their own characteristics. Several of them were tested and criticised.

Finally, even if we were to use a hypothetically perfect D-criterion, it would not be enough to prove the validity of the meteor shower studied. An actual meteor shower is defined by its parent body, in the sense that each meteoroid comes from the same parent body. However, D-criteria alone cannot prove that several meteors come from the same source, even if their orbits are really similar. Indeed, non-gravitational forces and close encounters play a crucial role in the evolution of meteoroid streams and can group together meteoroids that did not come from the same parent body. 

Thus, a complete search of a meteor shower should not only use D-criteria but should also perform two more studies: a statistical analysis to show that the observed group is larger than expected based on random associations and a dynamical analysis to prove formally the common origin of the meteors studied. Even though some papers discuss the parent body of the group of meteors to try and prove they found a meteor shower, this should not replace a proper dynamical analysis. Indeed, even if the parent body cannot be found because it has been destroyed, a common origin can still be demonstrated. Furthermore, many papers propose a parent body based on orbital similarities but do not demonstrate that particles ejected from this body in the past match the observed meteors at time of detection. Other papers \citep{Trigo-Rodriguez_al_2007, Trigo-Rodriguez_al_2015,Pena-Asensio_al_2023} integrate backwards the meteoroids and the parent body, comparing their orbit using a D-criterion, to verify that the orbit similarity holds over time. This strengthen the claim on the parentage but it relies on a D-criterion, which shows once again how important it is to study the advantages and drawbacks, and especially reliability, of each D-criterion.

Several studies have shown the limitations of D-criteria, and even of orbital similarity in general \citep{Shober_al_2025,Shober_2025,Pauls_Gladman_2005}, but no paper that we know of lists the various criteria and their critics. This paper aims to fill that gap by providing a first review of the D-criteria used in meteor science, along with critiques of each criterion. This paper is the first step towards a complete and exhaustive study of D-criteria, which is necessary to strengthen claims of meteor shower discoveries.

In the next section, we explain how we gathered the statistics presented throughout the paper concerning D-criteria and other methods in papers searching for meteor showers. We then examine D-functions, first some derived from the $D_{SH}$ of \citet{Southworth_Hawkins_1963}, and then others that present new ways to define D-functions. In Sect.~\ref{sec:threshold} we explore what thresholds are usually chosen for these D-functions. A quick, non-extensive review of the grouping algorithms to find meteor showers follows. Then, in Sect.~\ref{sec:tests}, we describe tests that were performed on D-criteria and their algorithms. Finally, we conclude and make suggestions for future studies.

\section{Frequency of D-criteria in shower identification studies}\label{sec:meth_stat}

To demonstrate how most D-criteria are used, we performed a quick statistical analysis of papers searching for meteor showers. The goal was to provide the reader with some statistics on the use of D-criteria and in what context they are used. The Meteor Data Center (MDC) of the International Astronomical Union produces a list of meteor showers that are considered established \citep{IAU_MDC}. Several parameters of the showers are provided, with the papers that proposed these parameters. At the moment, 110 established meteor showers are listed, each with several nominal radiants and orbital parameters depending on the referenced study. Very often, the papers listed describe the search of meteor showers in a large dataset and thus end up mentioning several meteor showers. We only took into account papers that were easily accessible (paper published in open access and referenced in ADS), which brings the number of papers analysed to 43. We then counted how many used analysis of the radiants and/or D-criteria (and if yes, which one) and/or statistical analysis and/or dynamical analysis to find meteor showers. Amongst these 43, only three did not employ any of these methods, as they were only a compilation of previous works results. So they were taken out of the final counts, which means 40 articles were used in the statistics presented throughout this paper (see~\ref{appendix} for the complete list). We are aware that we are working with a limited number of papers, but this should give the reader an initial idea of how often each method and D-criterion is used. These papers are only used to present statistics on D-criteria, but the literature review we present throughout this article is based on a much wider variety of papers.

For example, 80\% of the papers studied took into account the radiant in their search, but most of them combined the radiant with other information. Still, 32.5\% of the papers only used the radiants to propose new meteor showers, while several of them mention that this search can only be considered as a preliminary analysis. An analysis of the radiants methods would therefore also be of interest, but is not the goal of this paper. However, 62.5\% of papers studied use D-criteria to conduct the search, which shows the importance of this method in the meteor science community.

\section{The \texorpdfstring{$D_{SH}$}{DSH} and functions derived from it}\label{sec:orbelts}

The $D_{SH}$ by \citet{Southworth_Hawkins_1963} is one of the oldest and probably the most well-know D-function. Two more were developed based on this one: \citet{Drummond_1981} ($D_D$) and \citet{Jopek_1993} ($D_H$). They are all based on orbital elements and use $q$ the distance to perihelion, $e$ the eccentricity, $i$ the inclination, $\Omega$ the longitude of the ascending node, $\omega$ the argument of perihelion, the subscript refers to one of the objects being compared. Below are their definitions.

\begin{equation}
    \begin{aligned}
        \boldsymbol{D_{SH}}^2 & = (e_2-e_1)^2 + (q_2-q_1)^2 \\
         & + \left( 2 \sin  \frac{i_2-i_1}{2} \right) ^2 + \sin i_1 \sin i_2 \left( 2 \sin \frac{\Omega_1-\Omega_2}{2} \right) ^2 \\
         & + \left( 2 \frac{e_2-e_1}{2} \sin \frac{\Pi_{1,2}}{2} \right)^2, \\
         &\text{with:} \\
        \Pi_{1,2}  & = \omega_2 - \omega_1
        + 2 \arcsin \left( \cos \frac{i_1+i_2}{2} \sin \frac{\Omega_2 - \Omega_1}{2} \sec \frac{I_{1,2}}{2} \right), \\
        &\text{and:~} \\
        \left( 2 \sin \right. & \left. \frac{I_{1,2}}{2} \right)^2 = \left(2 \sin \frac{i_2-i_1}{2} \right)^2 + \sin i_1 \sin i_2 \left(2 \sin \frac{\Omega_2-\Omega_1}{2} \right)^2. \\
    \end{aligned}
\end{equation}

\begin{equation}
    \begin{aligned}
        \boldsymbol{D_D}^2 & = \left( \frac{e_1-e_2}{e_1+e_2} \right)^2 + \left( \frac{q_1-q_2}{q_1+q_2} \right)^2 + \left( \frac{I}{180^\circ} \right)^2 + \left( \frac{e_1+e_2}{2} \frac{\Theta}{180 ^\circ} \right)^2, \\
        &\text{with: } \\
        I & = \arccos ( \cos i_1 \cos i_2 + \sin i_1 \sin i_2 \cos (\Omega_1 - \Omega_2)), \\
        &\text{and:~}\\
        \Theta & = \arccos{( \sin \beta_1 \sin \beta_2 + \cos \beta_1 \cos \beta_2 \cos (\lambda_1-\lambda_2))}.
    \end{aligned}
\end{equation}

\begin{equation}
    \begin{aligned}
        \boldsymbol{D_H}^2 & = (e_2-e_1)^2 + \left( \frac{q_2-q_1}{q_2+q_1} \right)^2 + \left( 2 \sin \frac{I_{1,2}}{2} \right)^2 \\
        & + \left( \frac{e_1+e_2}{2} \right)^2 \left( 2 \sin \frac{\Pi_{1,2}}{2} \right)^2,
    \end{aligned}
\end{equation}
with $I_{1,2}$ and $\Pi_{1,2}$ defined the same way as $D_{SH}$. From comparing the three equations, it is quite clear that the last two functions are derived from the $D_{SH}$.

The $D_{SH}$ was created when databases of meteoroid orbits had already grown significantly in size. Interestingly, the orbits from the photographic database used by Southworth and Hawkins were seen at that time as very precise, which prevented some researchers from using this criterion for less precise radar measurements \citep{Kashcheyev_Lebedinets_1967}. This shows how this criterion was regarded as a big step forward at the time.

In our analysis of the papers cited in the MDC list of established meteor showers, we found that 47.5\% of them used the $D_{SH}$. This clearly shows how often this criterion is being used, even if the proportion tends to decrease in time. In contrast, the $D_D$ is only used in 17.5\% papers and the $D_H$ in 15\%. It is worth noting that these last two criteria are almost always used in conjunction: 4 papers (10\%) use the $D_{SH}$, the $D_D$ and the $D_H$ together and another one uses the $D_{SH}$ and the $D_D$ in complement of each other, with only two paper using the $D_D$ on its own and another one using the $D_H$ alone as well. It is also interesting that this practice of using several criteria together only appears after 2006, while $D_{SH}$ was always used alone before that date, even though the other criteria already existed for a while.

Each of the new D-functions discussed in this section tries to correct the previous one(s). \citet{Drummond_1981} makes several criticisms of the $D_{SH}$, starting with the units of each term. The second term is the only one with a dimension of distance, while the others are dimensionless. While it is very probable that the second term should be seen as divided by 1~au, which solves this problem, it is never stated outright in the original paper. Furthermore, Drummond shows that each term varies in a different interval, and criticises the use of chords to compute angles. His own criterion tries to solve these issues: the angles are computed without using the chord, each term is dimensionless and varies linearly between 0 and 1. However this choice of Drummond makes his criterion less accurate when the perihelion distance is small \citep{Jenniskens_2008}.

\citet{Jopek_1993} analyses and critics both D-functions to try and come up with a better one. He notes that Drummond gives more weight to some terms than to others: specifically the term in eccentricity greatly increases when the eccentricity is small. He also notes that, while the $D_{SH}$ has a physical definition, the $D_D$ does not. The $D_H$ is therefore defined as a mix between the two other functions.

Each of these D-functions got criticised for similar problems. The use of the longitude of the ascending node can be a problem, since this parameter varies a lot from one meteor to another in a singular meteor shower \citep{Steel_al_1991}. This is due to the duration of some meteor showers, which can lasts several weeks. More generally, time is often directly present in these functions, through the longitude of the ascending node or the solar longitude (used in all D-functions cited in this paper, except \citealp{Neslusan_2002} and \citealp{Jopek_al_2008}). A meteor group found with a D-criterion tends to last much less than the real meteor shower underlying it, because of the use of these elements. One solution was proposed by \citet{Steel_al_1991} and reused in \citet{Asher_Steel_1995}: taking out from the $D_{SH}$ the two terms in function of the longitude of the ascending node. To our knowledge, the reliability of this was never tested, but it can seem quite radical to simply take out these terms.

Another solution is to construct showers using several reference orbits that follow the evolution of the meteor shower \citep{Moorhead_2019}. This solution is pretty straight-forward: instead of comparing meteors together, at the risk of comparing meteors at the beginning and at the end of the shower, each meteor is compared to a set of orbits that represent the evolution of the shower in time. Of course, this technique is interesting when we already know the meteor shower, but would not work for detecting a new one.

Other remarks can be made on the physical aspects of these D-functions. The problem of orbit similarity is a four-variable problem, while all of these D-functions exist in a five-dimension space. This is because we are working within the framework of meteor showers, where the orbits of meteoroids all intersect the Earth, providing a constraint for the problem \citep{Valsecchi_al_1999}. This means the orbital elements can be linked together by the necessary close encounter of meteoroids with the Earth. Similarly, the observational errors of each orbital element are also related \citep{Jenniskens_2008}.

Another point to consider is the specifics of the database used to conduct the search. Both the $D_{SH}$ and the $D_D$ expect a specific behaviour in the errors of the
databases studied \citep{Brown_al_2008}. First, they were developed for photographic orbits, whose errors are smaller than radar measurements. Second, they also rely on the errors to be uniform in the whole dataset, which is probably rarely true (the weather, the observing geometry and the peak brightness, for example, are varying parameters that have an effect on the precision).

Finally, a mathematical analysis can also be made. The problem of orbit similarity would fit well with the mathematical definition of a distance (a function of two objects $x_1$ and $x_2$ that is always positive, reaches 0 only if $x_1 = x_2$, is symmetric, and satisfies the triangle inequality). However, neither the $D_{SH}$ nor the $D_D$ satisfy the triangle inequality, meaning none of them are distances \citep{Kholshevnikov_al_2016}. While we could propose that this does not negate the use of these D-criteria, the absence of a robust mathematical framework is a drawback, as it weakens the demonstration in the search of meteor groups. \citet{Kholshevnikov_al_2016} propose three other D-functions that are indeed distances. They are also defined using orbital elements, but they are very rarely used in the community. \cite{Jopek_2020} used it on near-Earth asteroids, but there does not seem to be any paper applying these D-functions to meteor databases, which means that these functions have not yet been tested in the framework of meteor showers.

This quick review of D-functions using orbital elements shows a couple of important points: none of them are without faults, although it seems the $D_H$ of \citet{Jopek_1993} is holding a little better under scrutiny. While the $D_{SH}$ has long been the reference in the domain, by virtue of being the first, simple to implement, and the most used, we do think it would be interesting to move to new, more robust, D-functions. This does not necessarily negate the work already produced using it. Because a lot of D-functions are defined such that their value aligns with D-functions values usually found in the literature, we may find similar results with other D-functions. But using D-function that are more robust would strengthen the claim made when defining a new meteor group.

\section{Other approaches to D-functions}\label{sec:noorbelts}

As orbital elements are derived from measured quantities, the errors can get quite large, and that is often the case for the semi-major axis (or perihelion distance) \citep{Egal_al_2017,Vida_al_2018}. For this reason, it is interesting to take a look at a D-function using observed quantities. This is what \citet{Valsecchi_al_1999} choose in the definition of their D-function $D_N$ and its simplified form $D_R$. They use $\lambda$ the solar longitude of the meteor (which we already showed might be a problem for long showers), $\boldsymbol{U} = (U_X, U_Y, U_Z)$ the geocentric unperturbed velocity, $U$ its module, and $\theta$ and $\phi$ two angles defining the speed orientation. The simplified $D_R$ is written in the case where only planetary secular perturbations affect meteoroid orbits (i.e. when the timescales are not too long and the close encounters are few enough or at least do not have a major effect).  Even then, it is considered by \citet{Valsecchi_al_1999} as a necessary but not sufficient condition to find a meteor group.


$\theta$ and $\phi$ are defined as such:
\begin{equation}
    \begin{aligned}
        \theta & = \arccos (U_Y/U), \\
        \phi & = \arccos (U_X/U_Z). \\
    \end{aligned}
\end{equation}

And the criteria themselves are defined in this way:
\begin{equation}
    \begin{aligned}
        \boldsymbol{D_N}^2 & = (U_2-U_1)^2 + w_1 (\cos \theta_2 - \cos \theta_1)^2 + \Delta \xi ^2, \\
        & \text{with: }\\
        & \Delta \xi ^2 = \min (w_2 \Delta\phi_I ^2+w_3 \Delta\lambda_I^2; w_2 \Delta\phi_{II}^2+w_3 \Delta \lambda_{II}^2), \\
        & \text{where:}\\
        & \Delta \phi_I = 2 \sin \frac{\phi_2-\phi_1}{2}, \\
        & \Delta \phi_{II} = 2 \sin \frac{180 ^\circ + \phi_2-\phi_1}{2}, \\
        & \Delta \lambda_I = 2 \sin \frac{\lambda_2-\lambda_1}{2}, \\
        & \Delta \lambda_{II} = 2 \sin \frac{180 ^\circ+\lambda_2-\lambda_1}{2}, \\
        \text{and: } \\
        \boldsymbol{D_R}^2 & = (U_2-U_1)^2 + w_1 (\cos \theta_2-\cos \theta_1)^2.
    \end{aligned}
\end{equation}

Both D-functions are however incomplete in their definition, as they both depend on weighting quantities $w_i$ that are not defined. More importantly, no method is given to choose their values. A possible solution is to simply choose all $w_i$ equals to one \citep[as it was done in][]{Jopek_al_1999}, but this solution lacks robustness. Indeed there is no proof that the D-function written with these values might not give the most accurate results compared to the same D-function with another choice of weighting elements.

One interesting proposition derived from \citet{Valsecchi_al_1999} is to directly use geocentric quantities, comparing them without using the D-functions \citep{Brown_al_2008}. Brown et al. compare the right ascension and declination of the radiant of the meteor, the geocentric speed and the time of meteor appearance, using a density map. This proposal highlights the interest in geocentric quantities, but it is clear that many users prefer the simplicity of a D-function that provides a single value over the complexity of comparing multiple values.

Another D-function using geocentric observed quantities is defined by \citet{Rudawska_al_2015}, from the solar longitude $\lambda$, declination $\delta$, right ascension $\alpha$ and geocentric speed $V_g$ of the meteors to be compared. This criterion is written as follows:

\begin{equation}
    \begin{aligned}
        D_X^2 & = w_{\lambda} \left( 2 \sin \frac{\lambda_1-\lambda_2}{2} \right)^2 \\
        & + w_{\alpha} (|V_{g,1}-V_{g,2}|+1) \left[ 2 \sin \left( \frac{\alpha_1-\alpha_2}{2} \cos \delta_1 \right) \right]^2 \\
        & + w_{\delta} (|V_{g,1} - V_{g,2}|+1) \left( 2 \sin \frac{|\delta_1-\delta_2|}{2} \right)^2 + w_V \left( \frac{|V_{g,1}-V_{g,2}|}{V_{g,1}} \right)^2.
    \end{aligned}
\end{equation}

Here, too, there is an issue with the weighting factors, as they are not defined. The definition of this criterion is also quite complex, making its use more daunting than other simpler criteria. Furthermore, the unit of the geocentric speed is not clear.

These last two D-functions are designed to compare the orbits of the meteors, even if they use geocentric elements to do so. Thus they should not be confused with a radiant comparison, which is often done very differently (usually through a graph representing the radiants, with the grouping often done visually). Both \citet{Valsecchi_al_1999} and \citet{Rudawska_al_2015} refer to the previously written D-functions and clearly use theirs in a D-criterion. This is why we include them both in our analysis.

Other D-functions were proposed, using orbital quantities other than orbital elements. \citet{Neslusan_2002}, for example, defines a criterion using only $\boldsymbol{c} = (c_x, c_y, c_z)$ the orbital momentum vector divided by the mass of the meteoroid: $C^2 = (c_{x1}-c_{x2})^2 + (c_{y1} - c_{y2})^2 + (c_{z1} - c_{z2})^2$. It should be noted that this criterion does not include any specifics linked to meteor showers.

\citet{Jopek_al_2008} put foward a new D-function, defined by the energy $E$, eccentricity vector $\boldsymbol{e} = (e_X, e_Y, e_Z)$ and $\boldsymbol{h} = (h_X, h_Y, h_Z)$ the angular momentum vector of the meteoroids compared, written as:

\begin{equation}
    \begin{aligned}
        D_V^2 & = w_{h,X} (h_{1,X}-h_{2,X})^2 + w_{h,Y} (h_{1,Y}-h_{2,Y})^2 \\
        & + 1.5 w_{h,Z} (h_{1,Z}-h_{2,Z})^2 + w_{e,X} (e_{1,X}-e_{2,X})^2 \\
        & + w_{e,Y} (e_{1,Y}-e_{2,Y})^2 + w_{e,Z} (e_{1,Z}-e_{2,Z})^2 + 2 w_E (E_1 - E_2)^2.
    \end{aligned}
\end{equation}

This time, the weighting factors are well-defined, as the standard deviation of the relevant vectorial elements of the orbits of known meteor showers. This is an interesting idea that could be applied to the previous D-functions, although, to our knowledge, no definitive study based on this concept currently exists.

As we explained before, \citet{Jenniskens_2008} notes that the orbital elements of a meteor are linked to its final destination, the Earth, which allows him to define a function $D_B$ based on orbital elements, written as:

\begin{equation}
    \begin{aligned}
        D_B^2 & = \left( \frac{C1_1-C1_2}{0,13} \right)^2 + \left( \frac{C2_1-C2_2}{0,06} \right)^2 + \left( \frac{C3_1-C3_2}{14.2^\circ} \right)^2, \\
        \text{with:}\\
        C1 & = (1-e^2) \cos^2 i, \\
        C2 & = e^2 (0,4 - \sin^2 i \sin^2 \omega), \\
        C3 & = \omega + \Omega.
    \end{aligned}
\end{equation}

In the case of $C3$, the value of $C3_1 - C3_2$ has to be the smallest difference between those angles. \citet{Jenniskens_2008} also defines another function, that is simply a comparison of $T_J$ the Tisserand parameter of the meteoroids: $D_T^2 = (T_{J,1}-T_{J,2})^2$. 

All of these D-functions are certainly interesting but there is clearly a lack of studies on this topic. Indeed we were not able to find papers that test these D-functions. Maybe the high number of existing D-function is the reason why most of them are not used. In our search in the MDC reference list, we only found two paper that cited the $D_N$ proposed by \citet{Valsecchi_al_1999} (one paper used it with $D_{SH}$ and $D_H$) and one that cites the criteria proposed by \citet{Jenniskens_2008}. While the literature as a whole seems to welcome more readily these new options, it seems meteor groups search is more often conducted with older criteria.

From previous criticisms, we would recommend the use of Valsecchi's function, despite the weighting problem, for its use of geocentric elements. No significant critics seem to have emerged against functions defined by \citet{Neslusan_2002}, \citet{Jopek_al_2008} and \citet{Jenniskens_2008}, which is sufficient to include them in the list of relevant criteria as well. However, it does seem that the function by \citet{Rudawska_al_2015} suffers from some lack of definition.

\section{Threshold values}\label{sec:threshold}

As explained in the Introduction, the test to know if two meteors might belong to the same group depends on the chosen threshold, which is in itself a difficult problem. Indeed, even if the chosen D-function was perfect, it would still fail if the chosen threshold is not adapted. Most papers simply use values that are found in the literature \citep{Drummond_1981, Jenniskens_2008}, which guarantees an homogeneity, but also means that several papers do not justify their choice.

However, most papers written on the subject tend to consider a threshold that varies according to the size of the database to analyse. Several formulas have been proposed \citep[for example]{Lindblad_1971_1}. \citet{Jopek_Bronikowska_2017} go one step further: their threshold depends not only on the size of the database but also on the D-function used. 

Choosing a threshold depending on the size of the database is not necessarily an obvious choice. Indeed \cite{Lindblad_1971_2} explain that the false positive rates in a bigger database can be higher only because they use a clustering algorithm that could form long chains of meteor. These chains are not realistic and should therefore be avoided (for a more detailed explanation, see next section on clustering algorithm). There is a higher chance for such chains to be produced with large database if the threshold is not appropriately constrained. However it is important to note that many papers choose their threshold according to the size of the database even when using a different type of clustering technique than the one described by \cite{Lindblad_1971_2}.

In an different direction, \citet{Jopek_al_1999} compute the threshold to make the search of meteor groups as accurate as possible. Here, the threshold is not a value that is adaptable to the method chosen, but part of the method itself. Their method uses datasets that are randomly generated following "the same marginal distribution in the variables as the real sample". They find the threshold such that, in r\% of cases, no group is found, with r the reliability level (99\% for example). One last step is to check that increasing the size of the generated datasets do not change much the value of the threshold. This threshold is then used in the search on the real dataset. The search used the $D_{SH}$ and Valsecchi's D-function, but it could be carried out with any D-function. This method ensure that the threshold is adapted to the dataset studied and reduce the possibility of detecting a group that is only a chance association. 

Finally, it is possible to not rely on thresholds and use a different approach, as shown by \citet{Neslusan_2002} who uses the "break-point" method. He draws a cumulative histogram of the number of meteors depending on the value of the D-function. A sharp turn should be observed, which marks the limits between the meteor group and sporadic meteors. However, this does not mean that all issues are solved by this method. A problem that may arise is that the sharp turn, which plays the role of the threshold in this method, depends on the proportion of the dataset that is composed of streams. If only a small part of the database is composed of streams, the threshold has to be low in order to be statistically significant; a higher threshold would mean a high number of false positives. 

These last two solutions seem very promising, in our opinion, as they tend to reduce the dependency on previous D-criteria and on previous meteor showers, and try to improve the accuracy of the search for meteor groups. It should be highlighted here that values of thresholds found in literature might not be reliable, because a good choice of threshold should always come from an estimate of the sporadic meteors background.

\section{Clustering algorithms using D-criteria}\label{sec:algo}
To search for a group of meteors (and eventually a meteor shower), a D-criterion has to be associated with an algorithm, that determine how the meteors should be grouped. This adds to the difficulty of evaluating the D-criteria available.

Although some algorithms were developed to function without a D-criterion \citep[see e.g.][]{Brown_al_2010,Svoren_al_2000}, we focus here exclusively on algorithms that use D-criteria

\citet{Lindblad_1971_1} uses $D_{SH}$, the only criterion existing at this time, to search for groups. The method he describes is the single-neighbour technique, which is still very widely used. The idea is to compare each meteor in pairs with all others in the database. If the D value is smaller than the threshold, they are considered as linked. Linking the meteors together results in the formation of groups. This technique was criticised because it may lead to long chains of meteors where each meteor is considered similar in term of D-criterion to another but where the extremities of the chain can be widely different. To raise the accuracy of this technique, a first step might be added: withdrawing meteors from the sample thanks to a D-function and a pretty lax threshold \citep{Jopek_al_2008}.

Another quite popular method is described by \citet{Sekanina_1970_1}. The idea is to form a group around a constantly updated mean orbit. While searching a database, supposedly including no known meteor shower, the initial orbit is simply a randomly picked orbit and the first group is built around it. When a first group is found, another orbit not belonging to the group is chosen as initial condition for the second group, and so on. To form a group, meteors are selected if they are close enough, in the sense of a D-criterion, to the initial orbit. Then a mean orbit is computed from the meteors chosen, weighting each orbit with the formula $1 - (D/D_S)$ with $D_S$ the threshold for the D-function and $D$ the value of the D-function. Then from this mean orbit, a new search is conducted, leading to the recomputation of the mean orbit. This iteration should stop only when the parameters are close enough between two iterations.

The main drawback of this method is the convergence of the iteration: it can be quite slow \citep{Sekanina_1970_1} or even not converge at all \citep{Jopek_Froeschlé_1997}. The method is not very efficient with low concentration data \citep{Arter_Williams_1997}. 

There seems to be some merit in combining two different D-criteria in the search. The idea is that the biases of one criterion can be compensated by the other. \citet{Rudawska_al_2015} explains how this can be performed. The procedure consists of two steps. First, a D-function with a strict threshold is used to conduct an initial search in the database, forming associations characterized by their mean elements. These associations are then regrouped if their mean elements are sufficiently close, according to another D-criterion (here, they use the $D_X$ they defined in the same paper). This second step must be iterated multiple times until reliable groups are formed. 

An alternative algorithm is defined by \citet{Welch_2001}, who uses a very different approach from the previous ones. The algorithm first creates a density map from the dataset, using a D-function as metric: when the meteors are close to the threshold of the D-criterion, the density rises. When the map is drawn, the peaks of density show where the heart of the groups are; they can be found with a simple gradient. The author explains how major and minor showers might be found this way and underlies the importance to check whether the peaks could be found randomly (this is achieved by applying the algorithm to an artificial database). It is also a method that employs a different use of the D-criterion, offering a distinct perspective. However, in their comparison of two D-criteria and several algorithms, \citet{Rudawska_Jopek_2010} show that the difference between this last method and the single-neighbour technique is not very significant. Therefore, it might not be worth using it, as it is more complex than a single-neighbour algorithm or even the algorithm described by \citet{Sekanina_1970_1}.

However, another density-based clustering algorithm proposed as an improvement is Density-Based Spatial Clustering of Applications with Noise (DBSCAN; originally formulated by \citealp{Ester_1996_DBSCAN}). DBSCAN searches for core meteors that have at least $N$ neighbours inside a hypersphere of radius $\epsilon$. All objects that are reachable from a core point are included within the group; objects that are not reached from any core are excluded. When $N =$\,1 the algorithm collapses to the traditional single-linkage (``single-neighbour'') chaining \citet{Lindblad_1971_1}. Raising $N$ suppresses the formation of unrealistic, elongated chains because every link now has to be supported by $N-1$ independent neighbours. DBSCAN therefore retains the inclusiveness of single-linkage, while adding two desirable properties. First, by insisting that every core meteor have at least $N$ neighbours inside the $\epsilon$-radius, it concentrates the group around a denser, more ‘central’ core and breaks up the long, fragile chains that single-linkage tends to produce. Second, any meteor that fails this density test is automatically labelled noise, so isolated sporadics sitting in low-density regions of orbital-element space can’t trigger or extend a cluster on their own. Additionally, the search only requires a range query on a kD-tree, the average complexity is $\mathcal{O}(n\log n)$, comparable to the pair-wise passes used in classical “friends-of-friends’’ methods but with a much lower false-positive rate when $N>$\,1.

\citet{Sugar_2017} were the first to adapt DBSCAN for the use of identifying meteor showers. \citet{Sugar_2017} adapted a similar methodology to that of \citet{Valsecchi_al_1999} by using geocentric parameters during the clustering ($\lambda_\odot$, $v_{g}$, $\beta_{g}$, $\lambda_{g}-\lambda_\odot$). They tested DBSCAN on Euclidean norm of the difference between the meteors' geocentric parameter vectors (see eq.1 in \citealp{Sugar_2017}), and applied it to 25\,885 fireballs observed by the NASA All-Sky Fireball Network and the Southern Ontario Meteor Network (SOMN). With $N=5$ and an $\epsilon$ chosen based on distribution of distances to the fourth-nearest neighbor because $N=5$, they recovered 25 strong and 6 weak meteor group in the combined NASA and SOMN data set. Subsequent shower and stream surveys have confirmed that DBSCAN performs well at effectively identifying clustering at minimal computational cost \citep{Shober_Vaubaillon_2024,Shober_al_2025,Ashimbekova_et_al_2025}. However, while there has been numerous studies done to estimate appropriate threshold for D-criteria using a ``single-neighbor'' algorithm (e.g., \citealp{Lindblad_1971_1,Jopek_Froeschlé_1997,Jopek_al_2008}), there is still no consensus on how best to set minimum neighbor to define a core point ($N$) for DBSCAN. In particular, the way that increasing or decreasing the minimum-neighbour requirement changes the statistical power of group detections, by trading off chain-breaking against completeness, has not yet been quantified. Another possible avenue to explore is a promising hierarchical extension of DBSCAN called HDBSCAN which builds a tree of clusters over a continuum of density thresholds, then extracts the most stable branches \citep{2017hdbscan}. HDBSCAN could be used to separate out a dense group core from its sparser filaments without the need to hand-pick an $\epsilon$. HDBSCAN was used recently in \cite{Pena-Asensio_Ferrari_2025}.

This short review does not mention all algorithms used, but it is enough to show that many exist, and combined with the large number of available D-criteria, this results in a very high number of techniques for forming meteor groups. This highlights the difficulty of testing all these combinations in order to provide at least a general idea of the accuracy of the available methods.

\section{Testing D-criteria }\label{sec:tests}

As stated before, 62.5\% of selected papers that search for meteor showers (see Sect.~\ref{sec:meth_stat}) use D-criteria to conduct their search. Another important figure is the number of papers that conducted a statistical analysis afterwards to check the obtained groups (5\%), and those that incorporated a statistical method in the search (20\%). This brings the number of papers using a statistical method at all to only 25\%, which is very low. 

Even more striking is the number of papers presenting an orbital analysis. Such an analysis should be based on dynamical elements, whether through a dynamical integration or through other means. Numerical simulations of that type seek to prove the common origin of the meteoroids in the stream, usually by integrating particles from that supposed point. A good example of such an analysis is \citet{Egal_al_2022} on the Taurids. In our sample of papers, only one discusses invariant orbital quantities, but this cannot be considered a dynamical analysis, as they do not check the validity of the stream itself. These figures show how the search is often thought to be valid when D-criteria are used, even when it is not corroborated by other methods. Additionally, 17.5\% of the referenced studies discuss the stream parenthood, usually trying to present only one parent body. 

Under these conditions, it is all the more important to make sure the D-criteria used are reliable. Most papers presenting a new D-function perform a test to show their validity \citep[with the exception of][]{Jopek_1993}. However, it should be noted that most of these tests are not designed to find the limits of these functions or to show in which case they fail, but rather to try and demonstrate how they can be used.

Most papers thus apply their criterion to a database of meteor data and try to find known meteor showers, as well as new meteor groups, often called "meteor showers" by the authors. This reasoning is observed in \citet{Drummond_1981}, for the criteria defined in \citet{Valsecchi_al_1999} (tested in \citet{Jopek_al_1999}), in \citet{Neslusan_2002} or in \citet{Jopek_al_2008}, where all authors compare their results with results obtained from the $D_{SH}$ (and even the $D_N$ in the last case).

\citet{Rudawska_al_2015} follow a similar procedure, but in this test, the criterion proposed is used together with the $D_{SH}$ to form the meteor group. While it can be interesting to combine several criteria in a search to maximise the accuracy, it does not seem apt to use a combination of criteria when trying to test one of them. Indeed, if some biases are found after testing, it would be impossible to attribute them to the $D_{SH}$ or to the $D_X$ independently, rendering the test useless.

\citet{Jenniskens_2008} applies his criteria to Near-Earth objects, which might not fit perfectly with our object of study, meteor showers. In a similar fashion, \citet{Kholshevnikov_al_2016} test their various criteria with two showers, but they seem to only take into account quite large bodies that may not be representative of the whole showers. This might be to ensure a better reliability of the orbits, but it still means these criteria are not tested on large meteor databases.

Only \citet{Southworth_Hawkins_1963} conducted a statistical test of their criterion, which, in our opinion, should show the path to follow for all D-criteria. Interestingly, in this case, the test was designed to evaluate the possibility of failure from this criterion, and not just to apply it. From the database used to find the meteor groups, they created an artificial database that followed the same distribution of sporadic meteors but that did not contain any groups. They then tested their criterion by applying it to this artificial database and recording the number of false positives. From this, they were able to deduce that half of the groups found thanks to the $D_{SH}$ could be formed randomly. However this number may be overestimated. First, \cite{Gartrell_Elford_1975} argue that no specifics in the distribution is taken into account. For example, because there are less meteors detected with a high inclination, a group formed with these kind of meteors is more trustworthy than the 50\% false positive rate implies. Another point we can add is the effect of the threshold. A higher threshold will lead to a higher false positive rate, while a lower threshold will diminish the false positive rate but also miss several meteor groups. The choice of threshold is thus a balancing act between a high false positive rate and a high false negative rate. Neither of the two articles mentioned in this paragraph discuss this. It is therefore highly probable that a better chosen threshold could improve the statistics given here.

D-criteria were also tested after their first definition, often comparing them together and trying to compute their accuracy. For example, a comparison between $D_{SH}$ and $D_V$ from \citet{Jopek_al_2008} was performed by \citet{Rudawska_Jopek_2010}. The procedure is as follows: a dataset of meteors containing both meteor showers and sporadics is generated (the showers are all ejected from near-Earth objects), and both criteria are used to find meteor showers in this dataset. Since the search for a meteor shower also depends on the grouping algorithm used (see section~\ref{sec:algo}), they applied several grouping algorithms for each criterion, reducing the risk of attributing a low score to a criterion because of the poor performance of the algorithm. 
They evaluated the performance of the criteria by counting the number of true and false positive, averaged by the number of meteor in the shower.
The entire procedure is repeated several times and a score is computed for each criterion/algorithm combination. Regardless of the conditions, the $D_V$ is always more reliable than the $D_{SH}$. Such a procedure is highly valuable and applying a similar method to all known criteria would be invaluable for understanding their reliability.

Statistics is a valuable tool for analyzing some problems, such as the probability of two similar orbits \citep{Pauls_Gladman_2005} or how to create a set of meteors that is both free of showers and statistically compatible with the actual set of meteors studied \citep{Jopek_Bronikowska_2017}. Using these results, it would be very interesting to evaluate the reliability of the D-criteria listed here.

A statistical analysis \citep{Pena-Asensio_Sanchez-Lozano_2024} was recently performed on the $D_{SH}$, $D_D$, $D_H$ and $\rho_2$ \citep[one of the D-functions developed by][]{Kholshevnikov_al_2016}, as well as a set of distance metrics, generally used with machine learning algorithms. The analysis uses the CAMS database to compare meteor clusters obtained thanks to the various methods tested with meteor showers from the IAU. Ranking of the methods are then provided following several statistical analysis. The paper concludes with a general appreciation of each method. In this analysis, the $D_{SH}$ does not perform worse than the other D-criteria, although $\rho_2$ performs even better. $D_D$ is considered the least accurate of all D-criteria. It should be noted however that the distance metrics performed better in most areas than the D-criteria tested. The authors therefore recommend using distance metrics above using D-criteria (personal correspondence with the corresponding author).

In line with \citet{Pauls_Gladman_2005}, \citet{Shober_al_2025} also present several tools that can be used to analyse in depth D-criteria, and apply them to the $D_{SH}$, the $D_D$ and the $D_H$. Those tools are both statistical and dynamical. Similar results were found when investigating the $D_N$ by \citet{Valsecchi_al_1999} in a similar manner \citep{Shober_2025}.

In terms of statistical analysis, Kernel Density Estimation (KDE) and probability functions are used to estimate the statistical significance of using D-criteria to form groups. KDEs are used to estimate the sporadic component within a specific population (NEO, fireball, etc.) and randomly draw from the estimated probability density function in order to constrain the expected similarity within a population due purely to chance. The KDE is advantageous because its estimation technique uses a kernel function whose bandwidth parameter provides an adjustable smoothing effect to the population. This can effectively remove small groups from the sporadic population estimate, however, if clusters or showers make up a large proportion of the dataset, they need to be manually removed \citep{Shober_Vaubaillon_2024}.

Several datasets are investigated in this way (meteorite falls, fireballs, USG impacts), and all of them appear to exhibit groupings more consistent with random association. Real groups might be found in larger databases, but these results should still caution us against using D-criteria outcomes without some statistical work to validate their relevance.

For orbital dynamics analysis, the decoherence lifetime was evaluated as the time it takes for meteoroids of a given group to drift too far apart to be recognized as similar by a chosen D-criterion. This is defined by \citet{Shober_al_2025} as the point at which more than 95\% of the original stream (referring to the number of particles, not the masses) is no longer included in the cluster. This is determined using a DBSCAN algorithm, where the epsilon parameter is set as the threshold for the $D_H$ criterion corresponding to the point where the cumulative similarity frequency curve changes for the population indicating an excess of similarity above background sporadic association levels (the "break-point" method as defined by \citealp{Neslusan_2002}). The number of minimum neighboring points to be defined as a "core point" can be modulated to increase and decrease the chain-like morphology of the groupings. Less neighbors required ($<$2) give long-chains and more neighbors required ($>$3) give more condensed groups that are less elongated. Lyapunov lifetime computations were also performed to evaluate chaos in those databases. 

The decoherence lifetime is of the order of tens to hundreds of thousand years, depending notably on the inclination, for Earth-crossing orbits (Fig.~15-17; \citealp{Shober_al_2025}). This number reaches 10 to 50 thousand years for essentially all asteroidal debris, visible as fireballs. The Lyapunov lifetime confirms the presence of chaos in meteoroid streams. Once more, these results highlight the risk of linking recklessly meteor groups and meteor showers: high chaoticity and low decoherence lifetime are both indicators that should warn us that a group of meteors might not be a meteor shower.

It is important to note that, irregardless of the D-function or criteria chosen or developed, there will never be a perfect identification method based solely on osculating orbital elements or impact conditions because streams are superimposed on a sporadic background that will always be inevitably unintentionally grouped in with true members. Even if the threshold is set to an astonishingly low value, there will always be a false positive rate. This rate, albeit, is characterizeable and should be minimized as much as possible without an equally unwanted high false-negative rate. To find a good happy-medium in this statistical balancing act, it is critical to understand exactly how poorly these criteria fail, and in which specific cases. Counter-measures could then be formulated. 

\section{Conclusion}\label{sec:ccl}

We have listed eight papers and the D-functions they present \citep{Southworth_Hawkins_1963,Drummond_1981,Jopek_1993,Valsecchi_al_1999,Rudawska_al_2015,Neslusan_2002,Jopek_al_2008,Jenniskens_2008}. The first three are variations of the $D_{SH}$ and are thoroughly commented on. However, all of these methods suffer from a lack of testing and verification. Only a few papers have applied statistical or dynamical methods to some of the D-criteria listed here, to assess where they succeed and, perhaps more importantly, where they fail. When such tests do exist, they highlight the need for caution when using any D-criterion.

This problem is complicated by two very important parameters needed when searching for meteor groups: the threshold of the D-criteria and the clustering algorithm. Both are often chosen based on the usual practices of the community, but there is also a lack of studies on which are the most reliable.

This paper only points towards the most lacking areas in our use of D-criteria. However, even without proper tests, one can restrict the number of D-criteria to be used. Indeed, the accuracy of a D-criterion is not the only parameter to take into account. A solid physical reasoning behind the criterion, a choice of variables with appropriate units, and a criterion specifically designed to account for meteors—these, in our opinion, provide sufficient reasons to favor some criteria. While testing all of them would be invaluable, strong and reliable proof for a meteor shower cannot be obtained if one uses a criterion that does not adhere to these guidelines.

An example of this reasoning would be the $D_{SH}$. While this function has been used a lot in the past and do not give worst results than the other orbital elements-based functions, its formulation has been questioned by several authors. While $D_{SH}$ was a tremendous step forward in its time, it was only a first step. Indeed, D-functions such as defined by \citet{Jopek_1993} seem more robust in their definition. We also would like to encourage the use of criteria by \citet{Valsecchi_al_1999}, \citet{Neslusan_2002}, \citet{Jopek_al_2008} or \citet{Jenniskens_2008}, as they all favour a novel approach on these topics. 

We have also pointed out which choice of threshold and algorithm seem the most reliable. In terms of thresholds, we recommend selecting it by maximizing the accuracy of the search \citep[using e.g.][]{Jopek_al_1999}. We also find the break-point method interesting, even if there are some precautions necessary to use it. In terms of algorithms, we recommend using multiple D-criteria to make the search more robust. The single-neighbor technique is a potential solution for linking meteoroids, but only if a preliminary search is conducted to avoid the 'long chain' problem. \citet{Sekanina_1970_1} algorithm can also be chosen, if its convergence is not an issue. Another robust algorithm seems to be DBSCAN, despite a lack of studies to choose some parameters. Indeed, it has been used in the context of meteors before and performs well.

Of course, no criterion is perfect, so the goal of this paper is not to advocate for the creation of more criteria, but rather to take a more thorough and comprehensive look at what we already have. Testing the existing D-functions, in combination with various threshold values and grouping algorithms under different conditions, is sure to yield a validated method for searching meteor groups across the many available databases. Thus we advocate for such tests to be performed. It is likely that, depending on the database, the preferred method will vary, which highlights the advantage of having a diverse range of criteria. 

However, this discussion should not obscure an important fact: even the best criterion alone is insufficient to confirm the validity of a meteor shower. This is well-established, as several MDC papers on established showers already mention it. Nonetheless, it is crucial to remember that meteor groups identified with D-criteria must also undergo statistical and dynamical testing. It is possible that our sample of papers does not fully reflect this reality, as the showers they discuss may have been tested in other studies not listed by the MDC. Indeed, the MDC tends to list papers that propose new parameters for showers, either in terms of orbits or observational elements. In any case, D-criteria should be viewed as just one step in the search for a meteor shower—an important step, to be sure, but not the final one.

In the future, we envision extensive testing of the D-criteria with dynamical analysis, which would quantify the reliability of these criteria.

\section*{CRedit authorship contribution statement}
\textbf{A. Courtot:} conceptualization, formal analysis, investigation, project administration, writing - original draft, writing - review \& editing. \textbf{P. Shober:} investigation, writing - original draft, writing - review \& editing. \textbf{J. Vaubaillon:} conceptualization, writing - review \& editing.

\section*{Declaration of competing interest}
The authors have no competing interest to declare.

\section*{Data availability}
No data was used for the research described in the article.

\section*{Acknowledgements}
We would like to thank Regina Rudawska for some exchanges about her criterion, and Eloy Peña-Senso for discussions about his article. We also thank warmly Marc Fouchard, who was supporting the very early stages of this research and who provided some comments on this paper.

\bibliographystyle{elsarticle-harv} 
\bibliography{biblio.bib}

\appendix
\section{Papers used to compute the statistics presented}\label{appendix}

43 easily-accessible papers were listed as resources for the list of established showers by the MDC. Three of them \citep{Cook_1973,Jenniskens_1995,Jenniskens_Gural_2011} were withdrawn from our count because they only consists of a compilation of previous results, and thus did not describe the methods used to find the groups listed.

The rest of the papers are presented in the Table~\ref{tab} with relevant information for the computation of the statistics presented in this paper. They are listed in chronological order because the evolution of the use of the methods described is also interesting.

\begin{table*}
    \caption{Papers used to compute statistical information given in the text, with their respective parameters.}
    \label{tab}
    \centering
    \begin{tabular}{c c c c c}
    \hline \hline
         Paper & Radiant & D-criteria & Statistics used & Dynamics discussed \\
         \hline
         \citet{Jacchia_1963} & \ding{52} & \ding{53} & \ding{53} & \ding{53} \\
         \citet{Nilsson_1964} & \ding{52} & His own & \ding{52} & \ding{53} \\
         \citet{Kashcheyev_Lebedinets_1967} & \ding{52} & \ding{53} & \ding{53} & Parent body discussion \\
         \citet{Kresak_Porubcan_1970} & \ding{52} & \ding{53} & \ding{53} & \ding{53} \\
         \citet{Lindblad_1971_1} & \ding{52} & $D_{SH}$ & \ding{53} & \ding{53} \\
         \citet{Lindblad_1971_2} & \ding{53} & $D_{SH}$ & \ding{53} & \ding{53} \\
         \citet{Sekanina_1973} & \ding{52} & $D_{SH}$ & \ding{53} & \ding{53} \\
         \citet{Gartrell_Elford_1975} & \ding{52} & $D_{SH}$ & \ding{53} & \ding{53} \\
         \citet{Sekanina_1976} & \ding{53} & $D_{SH}$ & Used in search & \ding{53} \\
         \citet{Wood_1988} & \ding{52} & \ding{53} & \ding{53} & Parent body discussion \\
         \citet{Terentjeva_1989} & \ding{52} & \ding{53} & \ding{53} & \ding{53} \\
         \citet{Lindblad_al_1993} & \ding{52} & $D_{SH}$ & \ding{53} & \ding{53} \\
         \citet{Gavajdova_1994} & \ding{52} & $D_{SH}$ & \ding{53} & \ding{53} \\
         \citet{deLignie_Betlem_1999} & \ding{52} & \ding{53} & \ding{53} & \ding{53} \\
         \citet{Ohtsuka_al_2001} & \ding{52} & $D_{SH}$ & \ding{53} & Parent body discussion \\
         \citet{Porubcan_Kornos_2002} & \ding{52} & $D_{SH}$ & \ding{53} & Parent body discussion \\
         \citet{Jopek_al_2003} & \ding{53} & $D_N$ & \ding{53} & \ding{53} \\
         \citet{Uehara_al_2006} & \ding{52} & $D_D$ & \ding{53} & \ding{53} \\
         \citet{Brown_al_2008} & \ding{52} & \ding{53} & Used in search & \ding{53} \\
         \citet{Molau_Rendtel_2009} & \ding{52} & \ding{53} & \ding{53} & \ding{53} \\
         \citet{SonotaCo_2009} & \ding{52} & \ding{53} & Used in search & \ding{53} \\
         \citet{Molau_Kac_2009} & \ding{52} & \ding{53} & \ding{53} & \ding{53} \\
         \citet{Brown_al_2010} & \ding{52} & $D_D$ & Used in search & Comparison of \\
         & & & & invariant orbital quantities \\
         \citet{Molau_al_2012} & \ding{52} & \ding{53} & \ding{53} & \ding{53} \\
         \citet{Ueda_2012} & \ding{52} & \ding{53} & \ding{53} & Parent body discussion \\
         \citet{Holman_Jenniskens_2012} & \ding{52} & $D_{SH}$, $D_D$ & \ding{53} & Parent body discussion \\
         \citet{Greaves_2012} & \ding{52} & $D_H$ & \ding{53} & \ding{53} \\
         \citet{Holman_Jenniskens_2013} & \ding{53} & $D_{SH}$, $D_D$, $D_H$ & Used in search & \ding{53} \\
         \citet{Molau_al_2013} & \ding{52} & \ding{53} & \ding{53} & \ding{53} \\
         \citet{Segon_al_2013} & \ding{52} & $D_{SH}$ & Used in search & \ding{53} \\
         \citet{Steakley_Jenniskens_2013} & \ding{52} & \citet{Jenniskens_2008} & \ding{53} & \ding{53} \\
         \citet{Andreic_al_2013} & \ding{53} & $D_{SH}$ & Used in search & \ding{53} \\
         \citet{Andreic_al_2014} & \ding{53} & $D_{SH}$, $D_D$, $D_H$ & \ding{53} & \ding{53} \\
         \citet{Molau_Kerr_2014} & \ding{52} & \ding{53} & \ding{53} & \ding{53} \\
         \citet{Kornos_al_2014} & \ding{52} & $D_{SH}$ & Used in search & \ding{53} \\
         \citet{Rudawska_Jenniskens_2014} & \ding{53} & $D_{SH}$ & \ding{53} & \ding{53} \\
         \citet{Jenniskens_al_2016} & \ding{53} & $D_{SH}$, $D_H$, $D_N$ & \ding{53} & Parent body discussion \\
         \citet{Roggemans_al_2020} & \ding{52} & $D_{SH}$, $D_D$, $D_H$ & \ding{53} & \ding{53} \\
         \citet{SonotaCo_al_2021} & \ding{52} & \ding{53} & \ding{52} & \ding{53} \\
         \citet{Roggemans_al_2023} & \ding{52} & $D_{SH}$, $D_D$, $D_H$ & \ding{53} & \ding{53} \\
         \hline
    \end{tabular}
\end{table*}

When the column "Radiant" is marked "\ding{52}", the groups are formed at least partially thanks to radiants. \citet{Sekanina_1976}, \citet{Andreic_al_2013}, \citet{Rudawska_Jenniskens_2014} and \citet{Jenniskens_al_2016} do mention radiants, or even discuss them, but do not base their search on these quantities. \citet{Ueda_2012} do not explicitly state that they use radiants to form their groups, but it is implied in the paper. In a similar way, \citet{Steakley_Jenniskens_2013} form their first groups without a D-criteria, and imply it is accomplished using radiants. 

\citet{Nilsson_1964} do mention the $D_{SH}$ as a reference but without using it in the search. Still it brings to 50\% the number of papers mentioning this criterion as relevant.

\citet{Jenniskens_al_2016} choose a method to try and avoid a random set of meteors, but there is still no statistical check to verify the groups found. \citet{SonotaCo_2009} use statistical analysis in their search only in the sense where they note biases. \citet{Molau_Rendtel_2009} do use a statistical method, but they use it in conjunction with radiants, which does not fit with what we are studying (they decide that the radiant has to be in the probability distribution for the meteor to be considered part of the group).

As mentioned in this work, no paper perform a true dynamical analysis. \citet{Lindblad_1971_1} do check the correlation between the groups and the orbital energy, but this is not an orbital analysis.

Several papers also mention that the results should be taken with a grain of salt. \citet{Molau_Kac_2009} and \citet{Rudawska_Jenniskens_2014} note that their results are preliminary, \citet{Molau_al_2012} call their groups "candidates" for meteor showers, \citet{Andreic_al_2014} remind their reader that D-criteria do not prove that their groups are not chance alignment and \citet{Roggemans_al_2020} note that D-criteria can be misleading and can lead to false positive in meteor groups.

\end{document}